\documentclass[5p,times,num,preprint]{elsarticle}

\usepackage{amssymb}
\usepackage{amsmath}
\usepackage{xcolor}
\usepackage{caption}
\usepackage{subcaption}
\usepackage{wrapfig}
\usepackage{booktabs}
\usepackage{multirow}
\usepackage{makecell}
\usepackage{changes}
\usepackage{algorithm}
\usepackage{algpseudocode}
\usepackage[hidelinks]{hyperref}

\usepackage{listings}
\definecolor{eclipseStrings}{RGB}{42,0.0,255}
\definecolor{eclipseKeywords}{RGB}{127,0,85}
\colorlet{numb}{magenta!60!black}
\lstdefinelanguage{json}{
    basicstyle=\scriptsize\ttfamily,
    commentstyle=\color{eclipseStrings}, 
    stringstyle=\color{eclipseKeywords}, 
    numberstyle=\scriptsize,
    stepnumber=1,
    numbersep=8pt,
    showstringspaces=false,
    breaklines=true,
    frame=lines,
    string=[s]{"}{"},
    comment=[l]{:\ "},
    morecomment=[l]{:"},
    literate=
        *{0}{{{\color{numb}0}}}{1}
         {1}{{{\color{numb}1}}}{1}
         {2}{{{\color{numb}2}}}{1}
         {3}{{{\color{numb}3}}}{1}
         {4}{{{\color{numb}4}}}{1}
         {5}{{{\color{numb}5}}}{1}
         {6}{{{\color{numb}6}}}{1}
         {7}{{{\color{numb}7}}}{1}
         {8}{{{\color{numb}8}}}{1}
         {9}{{{\color{numb}9}}}{1}
}

\journal{Future Generation Computing Systems}

\begin{document}

\begin{frontmatter}



\title{HP2C-DT: High-Precision High-Performance Computer-enabled Digital Twin} 

\author[bscaff]{E. Iraola\corref{cor1}}
\cortext[cor1]{Corresponding author email: eduardo.iraola@bsc.es}



\affiliation[bscaff]{organization={Barcelona Supercomputing Center},
            addressline={Plaça d'Eusebi Güell, 1-3, Les Corts}, 
            city={Barcelona},
            postcode={08034}, 
            state={Catalonia},
            country={Spain}}




\author[bscaff]{M. García-Lorenzo}
\author[bscaff]{F. Lordan-Gomis}
\author[upcaff]{F. Rossi}
\author[upcaff]{E. Prieto-Araujo}
\author[bscaff]{R. M. Badia}

\affiliation[upcaff]{organization={CITCEA, Universitat Politècnica de Catalunya},
            addressline={Av. Diagonal, 647, Les Corts}, 
            city={Barcelona},
            postcode={08028}, 
            state={Catalonia},
            country={Spain}}

\tnotetext[pubnote]{Article accepted in \emph{Future Generation Computer Systems}. Please cite as: \\E. Iraola, M. García-Lorenzo, F. Lordan-Gomis, F. Rossi, E. Prieto-Araujo, R.M. Badia,
\textit{HP2C-DT: High-Precision High-Performance Computer-enabled Digital Twin}, Future Generation Computer Systems,
2025,
108333,
ISSN 0167-739X,
\href{https://doi.org/10.1016/j.future.2025.108333}{doi.org/10.1016/j.future.2025.108333}.}

\begin{abstract}
Digital twins are transforming the way we monitor, analyze, and control physical systems, but designing architectures that balance real-time responsiveness with heavy computational demands remains a challenge. Cloud-based solutions often struggle with latency and resource constraints, while edge-based approaches lack the processing power for complex simulations and data-driven optimizations.

To address this problem, we propose the \textit{High-Precision High-Performance Computer-enabled Digital Twin} (HP2C-DT) reference architecture, which integrates High-Performance Computing (HPC) into the computing continuum. Unlike traditional setups that use HPC only for offline simulations, HP2C-DT makes it an active part of digital twin workflows, dynamically assigning tasks to edge, cloud, or HPC resources based on urgency and computational needs.

Furthermore, to bridge the gap between theory and practice, we introduce the HP2C-DT framework, a working implementation that uses COMPSs for seamless workload distribution across diverse infrastructures. We test it in a power grid use case, showing how it reduces communication bandwidth by an order of magnitude through edge-side data aggregation, improves response times by up to 2x via dynamic offloading, and maintains near-ideal strong scaling for compute-intensive workflows across a practical range of resources. These results demonstrate how an HPC-driven approach can push digital twins beyond their current limitations, making them smarter, faster, and more capable of handling real-world complexity.

\end{abstract}



\end{frontmatter}



\section{Introduction}

Digital twins have emerged as powerful tools to integrate the physical and digital worlds, enabling advanced monitoring, control, and decision-making. Although much research has focused on their simulation and visualization aspects, less attention has been paid to integrating the computational demands of high-fidelity modeling with the need for rapid decision support. This gap becomes particularly relevant in highly distributed systems that need both low-latency responses at the Edge and high computational power to operate.

Therefore, a significant challenge in digital twin architectures is bringing together fast response times with massive computing capacity. Many existing approaches rely on centralized cloud solutions, which can be inadequate for real-time applications due to network latency. On the other hand, purely edge-based solutions lack the computational power needed for complex simulations and large-scale optimizations. The present work addresses these issues by proposing a novel reference architecture for digital twins, the \textit{High-Precision High-Performance Computer-enabled Digital Twin} (HP2C-DT), and an accompanying process framework for its implementation. 


A key contribution of our reference architecture is bringing High-Performance Computing (HPC) into the loop, allowing the digital twin to take on heavy computational tasks like probabilistic analysis, synthetic data generation, and model training while still keeping the fast response times needed for real-time decisions at the Edge. HPC stands out from general cloud computing or private clusters by offering dedicated high-performance resources with specialized parallel processing and high-memory bandwidth. This is especially important when quick results are needed, like in short-term simulations that help drive immediate decisions. With HPC, the system can scale its resources as needed, ensuring that even demanding computations can be handled efficiently without slowing down critical operations.

The HP2C-DT reference architecture is designed with three primary focuses: Information Technologies/Operation Technologies (IT/OT) integration, computational infrastructure, and data management. The IT/OT integration dimension ensures that edge components can be deployed on diverse hardware and operate with minimal latency. From an infrastructure perspective, the system dynamically assigns computational tasks to edge, cloud, or HPC resources based on urgency and computational load. A relevant innovation is the adaptive orchestration of tasks, where time-sensitive computations execute immediately at the Edge, while less urgent workloads are scheduled asynchronously on available HPC or cloud resources. At the same time, in terms of data management and communication, HP2C-DT balances computation with data exchange, ensuring that messages and sensor updates are transmitted between edge and cloud nodes without overwhelming the network. The proposed architecture handles this through a combination of heartbeats to track node status and local aggregation of measurements.

The proposed reference architecture defines a blueprint for the core components, their interactions, and best practices to develop digital twin systems. Building on this foundation, this article introduces the HP2C-DT framework, which has been designed and implemented to bridge the gap between the reference architecture and real-world applications. The framework not only specifies the enabling technologies and provides detailed guidelines for their integration, but also includes a working implementation that simplifies the development of digital twins.

At its core, the HP2C-DT framework integrates edge computing, remote cloud components (cloud or on-premises), and HPC resources into a cohesive system. A key element enabling this integration is COMPSs~\cite{lordan2014}, which orchestrates parallel and distributed execution. Taking advantage of the function-driven approach of the reference architecture, each node—whether at the Edge or in the Cloud—runs a COMPSs agent that coordinates with peers to offload tasks dynamically. This offloading mechanism also includes HPC resources, which allow heavy computations to be assigned to high-performance computing clusters when needed.

A use case of HP2C-DT is in electrical grids, where edge nodes are distributed over vast geographical areas and operational requirements have evolved considerably with the increasing integration of distributed energy resources interfaced through power electronic converters. Traditional system operation, based on SCADA measurements transmitted to a centralized Transmission System Operator (TSO) control center, was adequate when the grid was dominated by large synchronous generators whose dynamics evolved slowly and were sufficiently captured by multi-second SCADA reporting intervals. The widespread deployment of converter-interfaced resources has introduced faster and more complex transient behaviors that require high-resolution and time-synchronized measurements. Phasor Measurement Units (PMUs), now widely installed within modern Wide Area Monitoring Systems, provide such measurements with reporting rates of ~60 samples per second. While these technologies enable real-time observability, real-time detection and mitigation of faults, instabilities, or oscillations require reducing end-to-end latency not only in measurement acquisition but also in decision-making and actuation. For this reason, recent paradigms in power-system monitoring and control advocate for distributed or edge-based computation, in which part of the processing is performed by computational units geographically closer to the measurement devices. By decentralizing computation, the architecture avoids unnecessary communication delays associated with routing all data to the central control room and distributing control signals back to the field, thereby enabling faster local assessment and timely control actions. Within this context, HP2C-DT supports a division of responsibilities in which latency-critical analyses and immediate control actions are performed at the edge, while the centralized computation  infrastructure remains responsible for computationally intensive tasks such as predictive maintenance, load and generation forecasting, and probabilistic risk assessment. Similar scenarios arise in industrial automation, where production line adjustments require millisecond-level reaction times while complex optimization models refine scheduling and resource allocation, and in autonomous transportation networks, where real-time navigation depends on local edge decisions supported by large-scale traffic simulations and data-driven route optimization. Building on the reference architecture and implementation framework, we develop a prototype of a digital twin for power grids to test HP2C-DT’s main features in a real-world scenario.

This article is structured as follows. Section~\ref{sec:related} provides a concise review of related work on digital twin architectures, their connection to the computing continuum, and their use of HPC resources. Section~\ref{sec:architecture} outlines the proposed reference software architecture, followed by Section~\ref{sec:architectural-principles}, which examines the most important features of the reference architecture in detail. Section~\ref{sec:framework} introduces the HP2C-DT framework, extending the reference architecture for implementation. Section~\ref{sec:experimental} presents an experimental use case in power systems, demonstrating the benefits of HP2C-DT. Section~\ref{sec:discussion} discusses the previous experiments and their results, and examines the scope of the HP2C-DT framework for digital twins within the computing continuum. Finally, Section~\ref{sec:conclusion} provides concluding remarks.

\section{Related work}
\label{sec:related}

Michael Grieves~\cite{grieves2014} first introduced the concept of a \textit{digital twin} in 2014 as a virtual representation of a physical product, rooted in the management of the product life cycle. This involves a continuous exchange of data: the physical object sends real-time information to its virtual counterpart, while simulations or optimizations performed on the virtual object influence the physical object. In 2017, Grieves expanded this concept by introducing \textit{Digital Twin Instances}, which are individual virtual models of physical products; \textit{Digital Twin Aggregates}, which combine multiple instances; and the \textit{Digital Twin Environment}, a space where digital twins interact in multiphysics simulations~\cite{grieves2017}.

A more recent definition, provided by the CIRP Encyclopedia of Production Engineering~\cite{stark2019}, describes a digital twin as a digital representation of a unique product, system, or service that captures its characteristics, properties, and behaviors throughout various phases of the life cycle. The digital twin operates through models, information, and data, and enables real-time monitoring, optimization, and decision-making.

Jones et al.~\cite{jones2020} further refined the concept of a digital twin by identifying its key elements. A \textit{physical object} or \textit{entity}, such as a machine or system, has a corresponding \textit{virtual object} or \textit{entity} that digitally represents it. These entities exist within \textit{physical and virtual environments}, with different levels of \textit{fidelity} affecting their accuracy. This fidelity can be measured through the number of parameters transferred between physical and virtual objects. The \textit{state} of a digital twin depends on the measured parameters, such as temperature or voltage, which are collected through \textit{metrology}. The digital twin also includes \textit{processes} for updating the physical or virtual model based on new data. The \textit{physical-to-virtual connection} captures real-world data through sensors, while the \textit{virtual-to-physical connection} applies insights from simulations to control real-world devices.

\subsection{Twinning}

Less emphasis has been placed on how digital twins contribute beyond visualization and monitoring—specifically, in decision-making and model refinement. In Jones et al.'s terms, these capabilities fall under twinning, which refers to the synchronization of physical and virtual instances through metrology (reading an entity’s state) and realization (modifying an entity’s state). The frequency of twinning varies by application, with high-speed synchronization being critical in fields like power grids.

However, Jones et al. detach twinning from virtual processes. Virtual processes provide computational tools such as optimization, prediction, simulation, and analysis, but these tools are usually not involved in (a) updating the virtual instance when discrepancies arise between expected and measured parameters or (b) determining how the digital twin should respond to such discrepancies in the physical system. Since synchronization relies on continuous data exchange and analysis, the interaction with these computational tools is essential. Yet, how often and in what ways virtual processes should influence synchronization remains unclear. 

\subsection{Edge computing perspective in digital twins}

Integrating digital twins with the computing continuum offers advantages for low-latency applications. Qi and Tao~\cite{qi2019} propose a hierarchical architecture that combines edge, fog, and cloud computing to optimize data processing across layers in digital twin-based shop floor environments. This architecture enables real-time control at the Edge, medium-latency data integration at the fog level, and large-scale analysis in the Cloud, and addresses challenges such as bandwidth constraints and latency. Building on this foundation, Groshev et al.~\cite{groshev2021} introduce the concept of Digital Twin as a Service (DTaaS), highlighting the role of edge computing, network function virtualization (NFV), and 5G technologies. Their study explains how edge computing minimizes end-to-end latency by offloading tasks closer to physical devices, while 5G provides ultra-reliable low-latency communication (URLLC) for real-time applications like remote control and visualization. Together, these studies show how the computing continuum meets the strict low-latency and high-reliability needs of digital twins in Industry 4.0.

\subsection{Cloud-computing and HPC}

Another important part of digital twins is how they handle heavier computational load, usually leaning on cloud computing for its flexibility and ease of access. This processing power is needed for running complex simulations, real-time data analysis, and predictive modeling, and allows digital twins to mirror and anticipate the behavior of physical systems. But, while the Cloud works well in many cases, it can fall short when extreme performance is needed. High-Performance Computing (HPC), on the other hand,  offers more power and efficiency and can handle the heaviest workloads with higher speed and accuracy~\cite{munhoz2023}, yet it is still not widely used for digital twins. This is why integrating HPC into digital twin architectures has gained traction as computational demands for high-fidelity modeling and predictive analytics grow. The Digital Twin Engine (DTE) proposed by the European Centre for Medium-Range Weather Forecasts (ECMWF)~\cite{geenen2024} exemplifies this trend, leveraging EuroHPC supercomputers to deploy climate digital twins for simulations and data handling. The DTE emphasizes code optimization for accelerator technologies like GPUs, providing efficient use of HPC resources while maintaining interoperability across distributed systems. Similarly, recent work by Ares de Parga et al.~\cite{parga2024} proposes a parallel Reduced Order Modeling (ROM) workflow using COMPSs~\cite{lordan2014, tejedor2017} on HPC clusters. In general, we see that many efforts focus on using HPC resources for highly complex, physics-based simulations, but there is still a lack of close integration of these resources into the digital twin life cycle from a software perspective.

\subsection{Orchestration of asynchronous workflows}

Task orchestration plays a key role in managing the diverse resource demands in this area. Nie et al.~\cite{nie2023} propose a multi-agent, cloud-edge orchestration framework that uses the digital twin and the Industrial Internet of Things (IIoT) concepts to improve distributed production control. Their approach combines cloud-based production models with self-adaptive strategies to handle dynamic exceptions and make better use of resources. However, while the framework works well for distributed manufacturing, it mainly focuses on static scheduling and does not explore real-time synchronization across distributed resources in depth. Similarly, Nguyen et al.~\cite{nguyen2022} introduce a digital twin orchestration framework with features such as federation, translation, brokering, and synchronization, demonstrating its use in smart city applications such as hotspot prediction and driving assistance. Their work highlights the benefits of cross-domain collaboration and federated learning for AI model sharing, but it remains mostly conceptual and does not address performance optimization in resource-limited environments or large-scale industrial settings. In contrast, Ares de Parga et al.~\cite{parga2024} take a more hands-on approach, evaluating the COMPSs framework for orchestrating complex workflows in heterogeneous environments. This way, the authors effectively manage task dependencies and resource allocation in distributed systems, making it a promising option for digital twin orchestration. 

\subsection{Gaps and novel contributions}

Existing architectures often isolate HPC, edge, and cloud resources, resulting in inefficiencies in workflows that demand both low-latency responses and intensive computation. Although domain-specific projects optimize HPC for modeling and simulation, they do not incorporate real-time decision-making at the Edge. Conversely, hyper-distributed platforms prioritize the edge-cloud balance but omit HPC for large-scale simulations. This article proposes a reference software architecture that addresses these gaps by introducing HPC in the loop within a parallel and distributed orchestration component, together with IT/OT integration at the Edge.

\section{HP2C-DT reference architecture}
\label{sec:architecture}

\begin{figure*}[t!]
    \centering
    \includegraphics[width=0.9\linewidth]{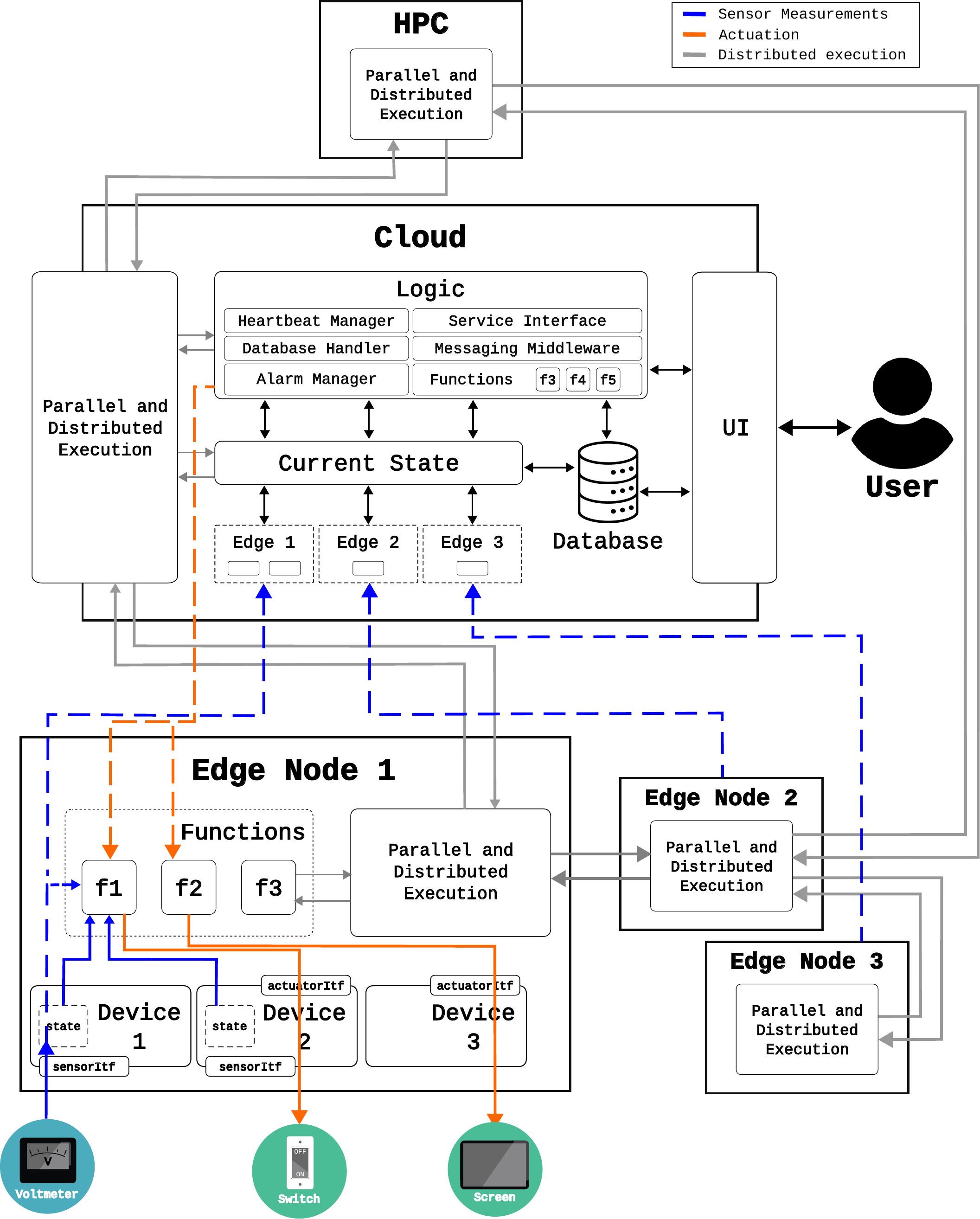}
    \caption{Reference architecture diagram.}
    \label{fig:arch-detail}
\end{figure*}

The HP2C-DT reference architecture is structured into three layers: Edge, Cloud, and HPC, as shown in Figure~\ref{fig:arch-detail}. The architecture is designed so that the Edge layer focuses on the underlying infrastructure, while the Cloud layer provides a global perspective of the digital twin. Meanwhile, the HPC layer supplies the raw computational power needed for intensive processing.

The Edge layer manages physical devices and local computing resources. It operates autonomously, processing information locally to meet real-time requirements. This layer handles immediate control tasks and real-time monitoring, ensuring quick responses to changes in the environment. For example, sensor measurements collected by edge nodes are transmitted to the Cloud layer using a publish-subscribe messaging model (\textit{sensor measurement} lines in Figure~\ref{fig:arch-detail}). At the same time, actuation commands from the Cloud are sent back to specific edge nodes through point-to-point messaging (\textit{actuation lines} in the figure). 

The decision between publish-subscribe and point-to-point communication arises from the asymmetry in control flow and communication patterns. Edge-to-Cloud communication mainly transmits sensor data upstream, which can be high-frequency, high-volume, and requires many-to-one ingestion. It must also scale with a growing number of edge producers. A publish-subscribe model suits these needs and allows loose coupling between producers and consumers while supporting expansion and modularity. In contrast, Cloud-to-Edge communication carries commands triggered by decisions at the Cloud layer, targeting specific actuations at the Edge. These messages are typically lower in frequency (since higher-frequency actuations are handled locally to enable faster response) and require precise delivery to the correct edge node. This fits a point-to-point communication model, which favors simplicity and determinism, allows each edge node to handle incoming control messages with node-specific logic, and supports bidirectional exchanges through acknowledgments or return responses. While other models such as direct APIs, message queues with explicit routing, or narrowly scoped publish-subscribe topics could technically support these patterns, they often introduce tighter coupling, higher complexity, or scaling overheads that our chosen model avoids in this architecture.

The Cloud layer acts as the brain of the system, integrating various components such as virtual representations of devices, database storage, computation tools, and user interfaces. It collects and processes information from multiple edge nodes, storing measurements in a time series database, and triggering alarms when needed. The Cloud layer also facilitates advanced computations like simulations and probabilistic analyses by leveraging its access to the full digital twin representation. Its messaging middleware manages communication with edge nodes, while other modules handle tasks like real-time monitoring (via heartbeat signals), alarm logging, and exposing APIs for external services.

Finally, the HPC layer provides high-performance computing capabilities for resource-intensive tasks that exceed the capacity of the Edge or Cloud layers. HPC systems consist of several computing nodes (generally servers with high computing power and a large amount of memory) interconnected with a high-speed network and designed for parallel processing of complex problems. These systems often incorporate specialized hardware accelerators---GPUs (Graphics Processing Units), FPGAs (Field-Programmable Gate Arrays), or TPUs (Tensor Processing Units)---to solve specific problems. Some cutting-edge HPC systems also integrate emerging technologies like quantum computing, which leverages quantum mechanics to solve certain types of problems exponentially faster than classical computers. 

HPC systems outperform the Cloud in several ways: they offer heterogeneous hardware better suited to application requirements, lower-latency and higher-bandwidth networking, and optimized software workflows without virtualization overhead. For organizations with consistent high-performance computing needs, in-house HPC can also be more cost-effective than indefinite cloud rentals. In addition, stricter user control in HPC environments enhances data privacy and security by keeping data within the organization and reducing exposure to external threats. Unlike cloud platforms, HPC systems restrict direct service deployment, instead relying on job schedulers like SLURM, which streamline access control and ensure exclusive resource allocation. However, this scheduling mechanism means executions may not start immediately if resources are unavailable.

Beyond the capabilities of each layer, the reference architecture orchestrates computation across them to balance real-time responsiveness and large-scale processing. Computation tasks are referred to as functions and are classified based on latency requirements: synchronous functions execute locally for immediate responses, while asynchronous functions are scheduled across edge, cloud, or HPC nodes depending on resource availability. The Parallel and Distributed Execution component, present on every node, manages this orchestration. For computationally intensive tasks such as large-scale simulations or optimizations, it offloads execution to the HPC layer.

\section{Key architectural principles}
\label{sec:architectural-principles}

The HP2C-DT reference architecture balances real-time responsiveness at the Edge with large-scale computational capabilities. This is achieved by integrating operational and computational layers to handle execution, communication, and data flow. This section details the core architectural principles that enable this integration. Implementation-specific details such as input/output definitions and execution parameters are described in Section~\ref{sec:framework}, where we show how these features are concretely defined and managed in the HP2C-DT framework.

\subsection{IT/OT integration}
\label{subsec:it-ot}

A key aspect of the proposed reference architecture is its interaction with the physical environment within a digital twin. For this, the architecture follows a taxonomy of digital twin entities that mirror physical and logical components. The Edge layer is responsible for interacting with \textit{physical objects}, collecting sensor data, and executing control actions on them. To achieve this, edge nodes maintain an image of each connected physical device—a \textit{physically bound digital object}—which keeps a record of the device data received and can communicate with it to apply actions when possible. In contrast, the \textit{virtual digital object} in the Cloud layer provides a more approximate image of the devices. This way, the Edge layer handles a smaller volume of information and remains close to the physical world, while the Cloud layer aggregates data from numerous edge nodes with a lower burden. Consequently, approximation and aggregation are required to manage scalability, reduce communication overhead, and extract higher-level insights from distributed edge data sources to the cloud node. Section~\ref{subsec:data} offers more information about such mechanisms.

Following this logic, each physical object in the digital twin is associated with a \textit{physically bound digital object} and a \textit{virtual digital object}. Within HP2C-DT, and from a software perspective, these are referred to as a \textit{device}. A device functions as a \textit{sensor} if it produces a measurement stream, an \textit{actuator} if it performs actions that modify the physical environment, or both if it supports both roles.

\begin{figure*}[!htbp]
     \centering
     \begin{subfigure}[b]{0.66\textwidth}
         \centering
         \includegraphics[height=9cm]{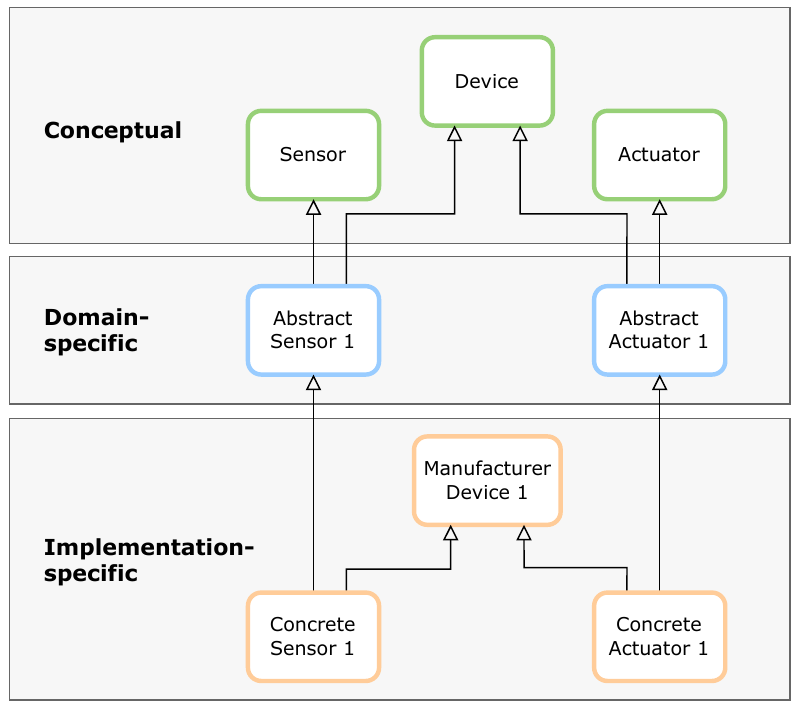}
         \caption{Levels of abstraction abstraction of digital twin objects.}
         \label{fig:y equals x}
     \end{subfigure}
     \hfill
     \begin{subfigure}[b]{0.32\textwidth}
         \centering
         \includegraphics[height=9cm]{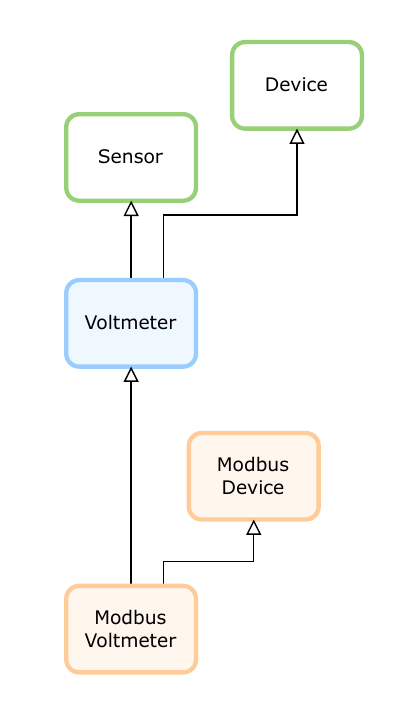}
         \caption{Voltmeter example.}
         \label{fig:three sin x}
     \end{subfigure}
     \caption{Hierarchical abstraction of digital twin objects.}
     \label{fig:device-hierarchy}
\end{figure*}

Figure~\ref{fig:device-hierarchy}a illustrates the hierarchical abstraction of devices. The first conceptual layer consists of the \texttt{Sensor}, \texttt{Device}, and \texttt{Actuator} abstractions, which define the core functionalities required for interaction within the architecture. All sensor-based devices must support triggering predefined sequences of actions when configured to do so, while actuator devices allow actions to be performed on the hardware when required. The second layer introduces domain-specific abstractions, such as \texttt{Abstract Sensor 1} and \texttt{Abstract Actuator 1}, which refine the base abstractions by specifying data types and measurement magnitudes. Finally, the implementation-specific layer contains concrete classes that inherit from the previous abstractions, integrating hardware- and manufacturer-dependent features. In the figure, \texttt{Concrete Sensor 1} extends \texttt{Abstract Sensor 1}, while \texttt{Concrete Actuator 1} extends \texttt{Abstract Actuator 1}. This layer defines low-level interactions with hardware, including communication protocols and data handling. Concrete devices also inherit from vendor-specific classes, such as \texttt{Manufacturer Device 1}, which encapsulates hardware drivers and communication details for a specific device family.

Consider the digital twin of a power grid. As represented in Figure~\ref{fig:device-hierarchy}b, the voltmeter sensor would be represented as an abstract class, \texttt{Voltmeter}, in the domain-specific layer, inheriting from the \texttt{Device} and \texttt{Sensor} abstractions. The \texttt{Device} abstraction handles properties such as labeling and positioning to ensure proper system integration, while the \texttt{Sensor} abstraction standardizes data collection and processing across different sensors. A concrete implementation of the voltmeter abstraction would extend this structure by integrating manufacturer-specific communication details. In the example of Figure~\ref{fig:device-hierarchy}b, the concrete class \texttt{Modbus Voltmeter} would implement the \texttt{Modbus Device} class, applicable in cases where multiple devices use the Modbus protocol~\cite{modbus}, a common industrial communication standard.

This hierarchical structure enhances flexibility and simplifies system extensibility. It enables a clear separation of concerns, from defining high-level functionalities to managing detailed hardware implementations. This approach also facilitates supporting devices with varying capabilities and those from different manufacturers, without overwhelming the system design. The taxonomy of digital twin objects, based on their proximity to the physical datum, along with the hierarchical structure of their software implementations shown in Figure~\ref{fig:device-hierarchy}, forms the basis for IT/OT integration. Together, they help map operational data from physical devices to computational processes in the digital twin.

\subsection{Data processing management}
\label{subsec:funcs}

Functions are modular computation tasks defined by the user in the HP2C-DT architecture. Depending on urgency requirements, they can be configured as synchronous or asynchronous. Because they share the same execution model, they can be used in a similar way at both edge and cloud nodes. Synchronous functions execute locally and immediately on the requesting node. Asynchronous functions, managed by the Parallel and Distributed Execution component, are scheduled and distributed across the digital twin infrastructure in a Function-as-a-Service (FaaS) manner, executing on available resources or offloading to peer nodes. They may consist of a single task or form complex workflows with multiple dependent or nested tasks, each assigned to a node that meets its constraints. 

In addition to being synchronous or asynchronous, functions can be triggered in different ways depending on system needs:

\begin{itemize}
    \item Frequency-based trigger (\texttt{onFrequency}): Executes periodically at a fixed interval.
    \item Event-based trigger (\texttt{onRead}): Executes when a new measurement arrives from a specified sensor. It can be configured to run after a set number of measurements instead of every individual reading.
    \item Change-based trigger (\texttt{onChange}): Executes only when a sensor's state changes, reducing redundant computations and communication overhead.
    \item On-start trigger (\texttt{onStart}): Executes once at application deployment, useful for persistent components like messaging listeners that handle actuation signals between edge and cloud nodes.
\end{itemize}

From a state-management perspective, functions in our architecture are designed to be stateless: they compute outputs only from their current inputs. State is encapsulated at the level of \texttt{Sensor} objects (see Section~\ref{subsec:it-ot}), which maintain rolling windows or histories of past measurements (see Section~\ref{subsec:data}). Functions can be triggered by changes in these sensor states, but they themselves do not persist intermediate data across invocations. This separation of concerns simplifies distribution and scaling, since functions can be executed uniformly across the different architecture layers without managing internal state.

\subsection{Network traffic reduction}
\label{subsec:data}

The Edge layer can reduce communication overhead through aggregation methods, which is crucial in high-frequency domains, such as power systems. Each physically bound digital sensor object maintains a rolling window that stores recent measurements. Each entry consists of a timestamp and one or more recorded values, depending on the sensor type. The window is updated continuously as new data arrives, discarding the oldest entries to keep a fixed size. This follows the standard sliding-window buffering model commonly used in time-series processing.

Rolling windows serve multiple purposes within the digital twin architecture. Functions can leverage them to compute results over recent time frames, but their primary role is to enable aggregation. Without aggregation, sending each measurement separately would overwhelm the network as the number of distributed nodes scales. Instead, edge nodes can transmit entire rolling windows---preserving all data while controlling transmission frequency---or apply aggregation methods that further reduce message size. The latter is particularly useful in environments with a high density of sensor information. Aggregations may include operations such as averaging, last-value reporting, or domain-specific reductions (e.g., phasor computations). The concrete implementation of these mechanisms is discussed in Section~\ref{sec:framework}.

\section{HP2C-DT framework}
\label{sec:framework}

The HP2C-DT framework is a full implementation of the HP2C-DT reference architecture, that provides a modular and extensible software solution to deploy digital twins that take advantage of edge, cloud, and HPC environments. It defines standardized components for device management, computation, communication, and data aggregation. This way, the framework enables developers to implement domain-specific digital twins while benefiting from a robust underlying infrastructure that simplifies integration. Through its modular design, it supports hardware-agnostic deployment, making it suitable for a wide range of industrial applications.

In this framework, edge and cloud nodes run as Java applications inside Docker containers~\cite{docker}. Java ensures platform independence, efficient concurrency, and access to a rich set of libraries, while the Docker container isolates dependencies and configurations, ensuring that each edge node runs in a controlled and reproducible environment free from compatibility issues. Each of these nodes contains an instance of a COMPSs agent~\cite{lordan2021} that implements the Parallel and Distributed Execution component of the reference architecture.

For communication, RabbitMQ~\cite{rabbitmq} is used as the messaging middleware, with a central broker running in the Cloud layer. This setup supports both publish-subscribe and point-to-point messaging, as required by the reference architecture. RabbitMQ offers message decoupling while also providing features such as queuing, delivery guarantees, and clustering. Its flexibility and open-source nature allow for partial decentralization, prevent vendor lock-in, and support the hybrid communication model defined by the HP2C-DT reference architecture.

The HP2C-DT framework is publicly available at \url{https://github.com/bsc-wdc/HP2C-DT} \cite{hp2cdt-zenodo}. The following sections go through the most significant features of the framework. It is important to emphasize that the examples used throughout the rest of the article illustrate the framework’s capabilities and potential, but do not represent a comprehensive view of all the functionalities required by a domain-specific, fully operational, end-to-end digital twin.

\subsection{COMPSs Agents}
COMPSs is a programming framework designed to boost developers' productivity working on parallel and distributed applications. By abstracting the complexity of parallel and distributed execution, COMPSs allows developers to focus on algorithm design without worrying about low-level details such as task synchronization, data transfers, or workload distribution. 

To do so, COMPSs follows a task-based approach. Programmers write sequential code in a plain major programming language---currently, it supports Java, Python, and C/C++---with no reference to any specific API to explicitly manage infrastructure and parallelism. By defining an interface or using decorators, programmers select a set of methods whose invocations will be replaced by an invocation to a runtime system that will handle its execution as an asynchronous task. The runtime system supporting the programming model analyzes the data accessed by each invocation to detect potential data dependencies with previously detected tasks and build a task-dependency graph. The runtime orchestrates the execution of all detected tasks across a distributed infrastructure, ensuring the sequential consistency of the original code.

The Colony framework~\cite{lordan2021} introduces an agent-based deployment strategy for COMPSs, allowing any node participating in the infrastructure to automatically convert sequential functions into task-based workflows. By deploying an agent on each node in the infrastructure—whether at the Edge or in the Cloud layers—the HP2C-DT framework facilitates this transformation, allowing functions (described in Section~\ref{subsec:funcs}) to be seamlessly converted into individual COMPSs tasks or orchestrated into structured workflows with appropriate resource allocation. This ensures that code execution fully leverages the available computing resources, including multiple CPU cores, GPUs, and other accelerators embedded in the node. In addition, the framework supports decentralized peer-to-peer orchestration of task executions. Agents collaborate to distribute the workload, allowing a function triggered on the Edge to offload all or part of its workflow to the Cloud layer or neighboring edge nodes. Each agent does this by checking its own resources to ensure that task scheduling stays within limits, and by evaluating the constraints of each task. For example, some tasks may require a specific processor architecture or a minimum number of CPUs. COMPSs agents handle this automatically. 

Likewise, COMPSs agents can be configured with HPC clusters as execution resources, making the HPC layer directly accessible from any node in the framework. This is enabled by a dedicated COMPSs connector for HPC clusters, which interfaces with the underlying batch management systems, handles job submission and tracking—including failure recovery—and manages result serialization and retrieval. Through this mechanism, COMPSs bridges the framework with HPC infrastructures, allowing cloud or edge nodes to offload intensive workloads without additional development effort on the HPC side.

\subsection{Edge layer implementation}

In the Edge layer, communications are implemented to ensure adequate data exchange between edge nodes and the Cloud layer. Each edge node runs a publisher function that sends messages to a RabbitMQ queue with the routing key \texttt{edge.<EDGE\_ID>.sensors.<SENSOR\_ID>}, which the server reads. Furthermore, a consumer function listens for incoming actuation commands on \texttt{edge.<EDGE\_ID>.actuators.<SENSOR\_ID>}.

Each edge node loads a JSON configuration file. Listing~\ref{lst:global-setup} shows part of a configuration file for an edge node labeled \texttt{edge1}. The \texttt{window-size} field sets the default rolling window size for all devices, though individual sensors can override it (see Section~\ref{subsec:framework-windows} for details on how the framework handles data management). The \texttt{comms} section configures communication protocols and hardware-specific parameters. The field \texttt{devices} lists the sensors and actuators. Section~\ref{subsec:framework-devices} provides more details about this setup. The field \texttt{funcs} specifies the computational functions that will run at the edge node, as explained in Section~\ref{subsec:framework-functions}.

\begin{lstlisting}[language=json, caption=Example of global configuration of an edge node (Snippet from setup file \texttt{edge1.json})., label=lst:global-setup]
{
  "global-properties": {
    "type": "edge",
    "label": "edge1",
    "window-size": 10,
    "comms":{
      "modbus": {
        "ip": "10.14.85.205",
        ...
      }
    }
  },
  "devices": [
    ...
  ],
  "funcs": [
    ...
  ]
}
\end{lstlisting}

\subsubsection{Device implementation hierarchy}
\label{subsec:framework-devices}

The HP2C-DT framework implements the \texttt{Device}, \texttt{Sensor}, and \texttt{Actuator} classes at the conceptual layer of the digital object hierarchy, following Figure~\ref{fig:device-hierarchy}. These come as a class for \texttt{Device} and interfaces for \texttt{Sensor} and \texttt{Actuator}, ensuring that any digital device can be implemented by extending them. The \texttt{Device} class handles instantiation from a JSON setup file, assigns labels and manages status availability. The \texttt{Sensor} and \texttt{Actuator} interfaces enforce the implementation of sensing and actuating methods. Specifically, \texttt{Sensor} requires a \texttt{sensed} method that dictates what happens when a new measurement arrives, including triggering computations and storing rolling windows. Similarly, \texttt{Actuator} defines how commands are received and executed. Developers can extend \texttt{Device} and implement \texttt{Sensor} or \texttt{Actuator} to support new measurement types while ensuring compatibility with the existing system.

The framework provides pre-implemented domain-specific classes that serve both as examples and as foundations for hardware-specific implementations. The current implementation focuses on power systems, offering abstract classes such as \texttt{Ammeter}, \texttt{Voltmeter}, \texttt{Wattmeter}, \texttt{Generator}, and \texttt{Switch}. The first three implement the \texttt{Sensor} interface and the last two, both \texttt{Sensor} and \texttt{Actuator}. Developers can use these as templates to create additional domain-specific classes. The built-in \texttt{MeasurementWindow}, defined on each of the abstract classes, maintains rolling windows of previous measurements, allowing historical data analysis (see Section~\ref{subsec:framework-windows}). In a similar way, concrete classes that handle communications with hardware are provided at the implementation-specific layer.

Listing~\ref{lst:device} shows a continuation of the edge configuration file in Listing~\ref{lst:global-setup} in which we set up a voltmeter object. The \texttt{label} field assigns a human-readable identifier, such as \texttt{Voltmeter Gen1}. The \texttt{driver} field specifies the concrete class, in this case, \texttt{ConcreteVoltmeter}, an implementation-specific component that ensures correct communication with the hardware on the Edge. The \texttt{properties} section defines operational parameters: \texttt{comm-type} specifies the Modbus protocol, \texttt{indexes} selects measurement channels, \texttt{window-size} determines the size of the rolling window (overriding the general window size set in global properties for this specific device, if set), and \texttt{aggregate} configures aggregation, here set to a phasor representation (see Section~\ref{subsec:framework-windows}). Users can specify multiple devices in the configuration file.

\begin{lstlisting}[language=json, caption=Example of configuration of devices (Snippet from setup file \texttt{edge1.json}), label=lst:device]
{
  "global-properties": {
    ...
  },
  "devices": [
    {
      "label": "Voltmeter Gen1",
      "driver": "es.bsc.hp2c.edge.manufacturer1.ConcreteVoltmeter",
      "properties": {
        "comm-type": "modbus",
        "indexes": [1],
        "window-size": 5,
        "aggregate": "phasor"
      }
    },
  ],
  "funcs": [
    ...
  ]
}
\end{lstlisting}

\subsubsection{Functions}
\label{subsec:framework-functions}

HP2C-DT functions can be implemented in Java, the architecture’s native language, or Python, which is supported through UNIX domain sockets for local execution. Python is included here to ease development and give access to scientific and AI libraries that are widely used in digital twin modeling and simulation. Each function follows a fixed signature, always receiving three arguments: a list or map of sensors it can access, a list or map of actuators it can control, and a map of additional \texttt{parameters} specific to the function’s requirements. This signature allows users to define various computational operations, such as model training, inference, state verification, or other custom computations. Additionally, the implemented triggers (see Section~\ref{subsec:funcs}) allow the execution of functions on demand. While HP2C-DT does not assign explicit priorities to triggers, each function is executed in a separate thread, relying on the operating system and JVM scheduler to manage concurrent execution. This approach has proven sufficient for typical workloads, and end users can implement custom prioritization or conflict resolution within their functions if needed.

In terms of interaction with the Parallel and Distributed Execution engine, which in this framework is implemented through COMPSs agents, asynchronous functions are handled in two complementary ways. At the intra-function level, functions are written as ordinary sequential code, and COMPSs transforms them into a workflow based on decorators and synchronization points added by the developer. This remains fully compatible with other parallelization mechanisms used inside the workflow itself, such as BLAS kernels invoked by NumPy’s linear algebra routines or OpenMP parallel regions that can be wrapped as COMPSs tasks. At the higher level, COMPSs schedules the tasks composing the workflows and offloads them across the available nodes. For that, it takes into account both the characteristics of the workload and the hardware resources of each node, which are described in the COMPSs configuration files.

Listing~\ref{lst:func-edge} shows the definition of functions in the \texttt{func} block through an example function, \texttt{VoltLimitation}, which triggers an action to disconnect the edge node from the grid when overvoltage is detected. The \texttt{lang} field specifies that this function is written in Java. The entry \texttt{type} specifies whether the function executes immediately or can be deferred. Here, the value \texttt{synchronous} indicates that the function's computation is urgent and must be performed as soon as it arrives. Otherwise, for the \texttt{asynchronous} cases, the execution would not be immediate and could even be offloaded to other nodes in the architecture, which would, in turn, be handled by the COMPSs agent. 

The \texttt{method-name} field provides the fully qualified class path where the logic of the function lies. The \texttt{parameters} section defines the function’s inputs and outputs. In this example, the function processes data from the \texttt{Voltmeter Gen1} sensor and controls the \texttt{Three-Phase Switch Gen1} actuator to disconnect the system. Furthermore, an additional parameter, \texttt{threshold}, is listed in \texttt{other}, indicating that the function enforces a voltage limit of 40~kV. The \texttt{trigger} section specifies that this function operates in an event-driven manner, using the \texttt{onRead} mechanism to execute upon receiving new sensor data (refer to Section~\ref{subsec:funcs} for information about the available function-triggering modes). The \texttt{trigger-sensor} field ensures that only readings from \texttt{Voltmeter Gen1} initiate execution, while the \texttt{interval} parameter configures the function to activate every five measurements instead of every single reading.

\begin{lstlisting}[language=json, caption=Configuration of functions (Snippet from setup file \texttt{edge1.json})., label=lst:func-edge]
{
  "global-properties": {
    ...
  },
  "devices": [
    ...
  ],
  "funcs": [
    {
      "label": "VoltLimitation",
      "lang": "Java",
      "type": "synchronous",
      "method-name": "es.bsc.hp2c.edge.funcs.VoltLimitation",
      "parameters": {
        "sensors": ["Voltmeter Gen1"],
        "actuators": ["Three-Phase Switch Gen1"],
        "other": {
          "threshold": 40,000
        }
      },
      "trigger": {
        "type": "onRead",
        "parameters": {
          "trigger-sensor": ["Voltmeter Gen1"],
          "interval": 5
        }
      }
    },
  ]
}
\end{lstlisting}

\subsubsection{Aggregation and rolling windows}  
\label{subsec:framework-windows}  

For data management, the HP2C-DT framework uses rolling windows that store measurements in each device object. The rolling window is implemented as a circular buffer: when it reaches full capacity, the oldest measurement is automatically replaced by the newest one. This design ensures constant-time insertions and deletions, making it optimal in both time and memory efficiency. Since no memory reallocation or element shifting is required, access to recent data remains fast.

The framework also provides implementations of aggregation methods, as seen in Section~\ref{subsec:data}, specifically the domain-agnostic ones: \texttt{all}, \texttt{average}, \texttt{sum}, and \texttt{last}. The \texttt{all} method is the trivial case where every measurement in the window is sent. \texttt{sum} transmits the sum of all values in the window, \texttt{average} computes and sends the mean, and \texttt{last} retains only the most recent value. We include both \texttt{sum} and \texttt{average} aggregators, which are mathematically closely related, because they are canonical examples of aggregation methods and are commonly supported in practical frameworks. Aggregation methods are easily configurable in the edge setup file, as indicated by the \texttt{aggregate} field in Listing~\ref{lst:device}. In addition, domain-specific aggregation methods may be useful, and the framework’s modular design allows them to be easily integrated.

A particularly valuable implementation is the phasor representation~\cite{ledwich2021}, which provides a compact representation of oscillatory signals. Phasor aggregation represents a periodic signal through a single complex number that encodes amplitude and phase. Given a sinusoidal signal with amplitude \( A \), angular frequency \( \omega \), and initial phase \( \phi \),  
\begin{equation}  
    x(t) = A \cos(\omega t + \phi)  
\end{equation}  
its phasor representation is $X = A e^{j\phi}$ or $X = A \angle \phi$. The frequency is omitted from the phasor representation because a nominal base frequency is assumed to be known and constant, as is standard practice in, e.g., power system analysis. This means that a steady-state sinusoidal signal $x(t)$ with known frequency can be represented by only two parameters, its amplitude $A$ and its angle $\phi$. We implement the \texttt{phasor} aggregation method using a Discrete Fourier Transform (DFT) approach to process the time series windows. The expected frequency of the signal needs to be provided beforehand in the function configuration.

Data aggregation induces information loss, which in turn can be assessed using the Mean Squared Error (MSE), $\varepsilon_{\text{MSE}}$, a standard metric for signal reconstruction fidelity. For a sensor producing a single value per measurement, given a rolling window \( X = \{ x_1, x_2, \dots, x_n \} \) of size \( n \), the information loss introduced by an aggregation function is:
\begin{equation}
    \varepsilon_{\text{MSE}} = \mathbb{E} \left[ (X - \hat{X})^2 \right]
    \label{eq:mse_loss}
\end{equation}
where \( \hat{X} \) represents the estimated reconstruction of \( X \) from the aggregated value \( Y = f(X) \).

For average aggregation, the estimate is simply \( \hat{x}_i = Y = \frac{1}{n} \sum x_i \), leading to:
\begin{equation}
    \varepsilon_{\text{avg}} = \left( 1 - \frac{1}{n} \right) \sigma_X^2
    \label{eq:avg_loss}
\end{equation}
which shows that for large \( n \), nearly all variance is lost.

For last-value aggregation, all previous values are ignored. Assuming an autoregressive process with correlation \( \rho \) between consecutive samples~\cite{brockwell1991}, the reconstruction error is:
\begin{equation}
    \varepsilon_{\text{last}} = (1 - \rho^2) \sigma_X^2
    \label{eq:last_loss}
\end{equation}
The autoregressive correlation $\rho$ is a model of signal smoothness: if consecutive measurements are highly correlated, keeping only the last value retains most of the signal’s information. If \( \rho \approx 1 \), little information is lost, but for uncorrelated data (\( \rho \approx 0 \)), most information is discarded.

The phasor transformation retains full information for pure sinusoidal signals if the estimated angular frequency, $\omega_{\text{est}}$, aligns with the actual angular frequency, ${\omega}$: 
\begin{equation}  
    \varepsilon_{\text{phasor}} \approx 0, \: \text{if } x(t) \text{ is a steady-state sinusoid and } \omega_{\text{est}} \approx \omega
    \label{eq:phasor_loss}  
\end{equation}  

For signals with harmonics or noise, phasor aggregation filters out these components, leading to an effective information loss of:  
\begin{equation}  
    \varepsilon_{\text{phasor}} \approx 1 - \frac{P_{\text{fundamental}}}{P_{\text{total}}}  
    \label{eq:phasor_loss_general}  
\end{equation}  
where \( P_{\text{fundamental}} \) is the power of the fundamental frequency component and \( P_{\text{total}} \) is the total signal power. This expression measures the fraction of signal energy that the phasor is able to retain when reflecting only the fundamental frequency, which is sufficient for many steady-state applications~\cite{ledwich2021}.

This analysis shows that information loss increases with \( n \) for sum and average aggregation, making them useful for noise smoothing but unsuitable for precise real-time tracking. Last-value aggregation, however, is more effective for highly correlated signals, as it retains the most recent relevant information with minimal error. Domain-specific methods such as the phasor representation can be particularly effective. In these cases, aggregation drastically reduces the size of messages while preserving essential signal characteristics, making it a practical choice for data aggregation.

\subsection{Cloud layer implementation}

The Cloud layer maintains a collection of virtual digital objects, each representing a physical device linked to an edge node. This collection updates dynamically with every incoming heartbeat and measurement, both of which are managed by the main RabbitMQ broker also running alongside the Cloud. The broker also handles point-to-point actuation d

To reduce communications beyond data aggregation, the Cloud can perform internal simulations to estimate device states when data is missing, delayed, or intentionally reduced to optimize communication. A two-sided simulation mechanism allows both the Edge and Cloud layers to estimate states independently, triggering data exchange only when discrepancies between estimations and actual measurements arise.

For time series data management, the Cloud layer integrates InfluxDB~\cite{influxdb}, optimized for fast ingestion and retrieval of sensor data, supporting downsampling and retention policies. The Cloud hosts an instance of this database, along with a dedicated database handler with utility methods for querying and writing data.

Monitoring, visualization, and user interaction rely on Django \cite{django} and Grafana~\cite{grafana}. Django serves as the web interface, allowing operators to monitor the status of edge nodes, track alarms, and perform manual actuation on distributed nodes. It also manages user authentication and integrates with the central database to display system updates in real-time. Grafana complements Django by embedding dynamic monitoring panels directly into the interface, offering real-time visualization of trends, anomalies, and system behavior. To facilitate external interactions, the Cloud layer exposes a REST API, enabling services such as real-time data sharing, status retrieval, and edge node configuration management.

Finally, the Server layer uses a JSON configuration file similar to that in edge nodes, but without a \texttt{devices} field. Instead, its knowledge of virtual digital objects is derived dynamically from edge node heartbeats rather than direct physical connections.

\section{Experimental evaluation}
\label{sec:experimental}

\subsection{Case study}
\label{sec: case_description}
\begin{figure*}[!htbp] 
    \centering
    \includegraphics[width=0.8\linewidth]{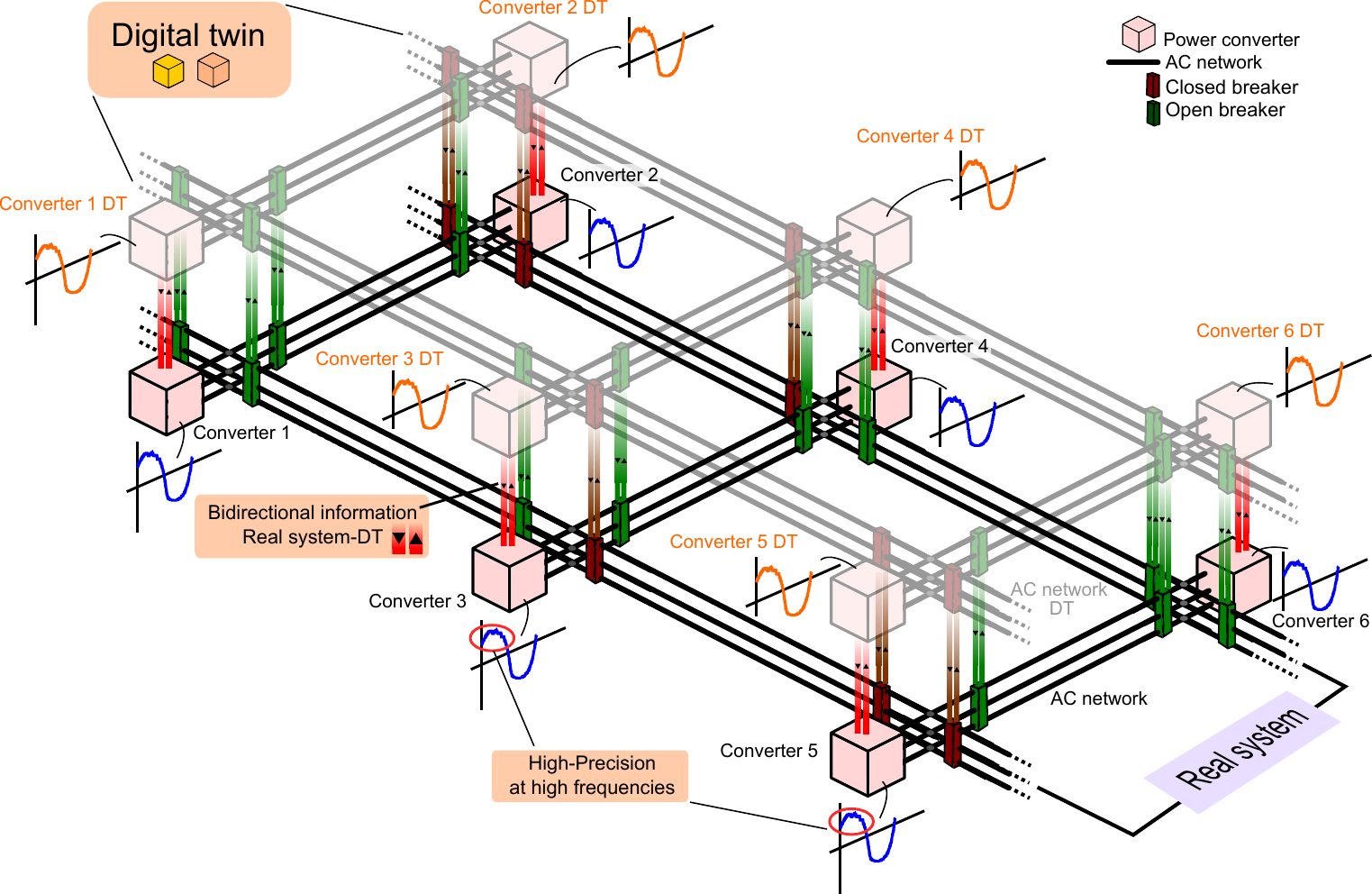}
    \caption{Power grid digital twin concept diagram.}
    \label{fig:power-grid}
\end{figure*}

Electric system operators currently use models that capture the actual state of their networks through Supervisory Control and Data Acquisition (SCADA) systems, allowing human operators to make informed decisions from a centralized control room. To make such decisions, operators mainly rely on the current state of the network, along with studies conducted using static (e.g., power flow) and dynamic (e.g., phasor-based) models extracted from the SCADA system.

This centralized approach has several limitations, including low-bandwidth decision-making, limited automation, and restricted control at the Edge. At the same time, the rise of renewable energy is shifting power systems from a centralized, controllable setup to a more variable and decentralized one. Addressing these limitations requires a more distributed and high-performance approach, which the HP2C-DT framework provides in the form of a digital twin architecture. It supports the integration of HPC resources and external Electromagnetic Transient (EMT) solvers, enabling EMT simulations while still providing access to traditional power-flow and phasor-based models, thereby ensuring multiple levels of modeling fidelity. It also enables computations at the Edge, where key assets of power systems—such as generators, power converters, transformers, protection relays, and metering stations—operate as HP2C-DT devices with their corresponding digital replicas. 

In the domain of power systems, digital twins are implemented as dynamic, real-time replicas of power system assets, supported by connection, data, model, and service layers. Previous efforts in this area typically do not address edge-to-cloud latency limitations, fast reaction, or the integration of HPC resources in the architecture; instead, they emphasize combining machine learning with simulation models for advanced analytics and feedback loops, where physical data refines virtual models, and tackle the challenges of data scarcity, dynamic topology changes, and uncertainties that hinder AI model performance~\cite{rahmani-saneApplicationsArtificialIntelligence2025}.

\subsection{Experimental setup}

We use a real-time simulator to model the power grid under different operating scenarios. The OPAL-RT hardware platform, OP4512,~\cite{op4512} runs a real-time EMT model of the electrical network using the Hypersim software~\cite{hypersim}, capturing dynamics in the kilohertz range. This is essential for accurately modeling power systems dominated by renewable energy. In our prototype, the OPAL-RT hardware represents the actual electrical grid and provides input and output interfaces. We create the necessary digital objects of the implementation-specific layer to interact with it, as explained in Section~\ref{subsec:framework-devices}.

Edge nodes run on a workstation directly connected to the OPAL-RT, ensuring real-time interaction. The workstation runs Ubuntu 22.04 and is equipped with an Intel i7-1265U CPU and 16 GB of RAM. Communication with the Cloud layer takes place over the institutional network using standard TCP/IP, with RabbitMQ as the messaging middleware. The workstation connects to the local network via Wi-Fi. The Cloud layer consists of virtual machines hosted on an on-premise server at BSC, each with 4~CPUs and 8~GB of RAM. Measured average round-trip latency between edge and cloud nodes is approximately 30 ms. HPC resources are provided by the general-purpose partition of the MareNostrum~5 supercomputer, which consists of compute nodes equipped with two Intel Xeon Platinum 8480+ processors—each with 56~cores at 2.0~GHz—for a total of 112 cores per node. Each node includes 512~GB of RAM and is interconnected via NVIDIA HDR InfiniBand for high-throughput, low-latency communication.

\subsection{Experiments}

In this section, we evaluate the system's performance through experiments that measure (1) communication bandwidth, (2) response time, and (3) HPC scalability. For this, we use the experimental setup described above. In particular, for (3), we select an application representative of a digital twin of a power system: stability analysis, and a standard case study, the IEEE 118-bus transmission model~\cite{pena2018}.

Stability analysis helps adjust the operation point of a power grid to specific operational objectives, such as reducing power generation or maximizing renewable energy consumption, without risking system stability. To enable stability analysis for real-time decision-making, we can leverage machine learning (ML) models, which can be trained to serve as surrogate models. In power systems, ML-based surrogates can replace stability assessment tools and allow stability evaluations to be carried out online on edge and cloud nodes~\cite{fajemisin2024optimization}.

However, training such ML models requires data. Although the digital twin continuously collects measurements from the operation of the actual system, such data may be insufficient to train accurate ML-based surrogate models, which requires the creation of a synthetic dataset \cite{thams2019efficient}. We can generate the synthetic dataset offline by exploring the operational space and applying conventional tools for power system stability assessment at each point. The HP2C-DT framework is well suited for this, as data exploration and generation are highly parallelizable and the framework integrates HPC throughout the digital twin’s architecture. Due to this, we propose a data generator tool to evaluate computation scalability on our prototype.

As shown in Algorithm~\ref{alg:entropy_exploration}, the exploration proceeds by sampling the entire operational space and progressively focuses on regions of higher interest---those intersecting the stability margin of the system---using an entropy-based criterion~\cite{rossi2022data}. After this, the tool recursively divides the variable space into subregions by segmenting the dimensions with a higher impact on system stability. Execution parameters include search space range, number of samples per subregion, and depth of exploration. The variable space consists of the power injected by each generator, the power demanded by each load, and several converter control parameters. The tool is implemented in Python using the VeraGrid library (previously known as GridCal~\cite{santiago_penate_vera_2025_14615064}), for feasible operating point power flow calculation, and the Python version of STAMP~\cite{arevalo2025matlab}, for an EMT-focused small-signal stability analysis.

\begin{algorithm*}[t!]
\caption{Entropy-based Recursive Subregion Exploration}
\label{alg:entropy_exploration}
\begin{algorithmic}[1]
\Require Dimensions of the search space and their ranges, number of samples per cell \textit{N}, objective function \textsc{Eval\_Stability}, maximum depth \textit{d}, grid configuration, use\_sensitivity
\Ensure Generated dataset with power flow results and stability labels
\State root\_cell $\gets$ entire exploration space
\State depth $\gets 0$
\State entropy\_parent $\gets 0$
\State dataset $\gets$ \texttt{empty}
\State \textsc{Explore\_Cell}(root\_cell, depth, entropy\_parent)
\Function{Explore\_Cell}{cell, depth, entropy\_parent}
    \State samples $\gets$ \textsc{Latin\_Hypercube}(cell, \textit{N})
    \State results $\gets$ \texttt{empty}
    \ForAll{sample in samples} \textbf{parallel}
        \State result $\gets$ \texttt{Eval\_Stability}(sample, grid configuration)
        \State results.append(result)
    \EndFor
    \State dataset.append(results)
    \State entropy\_child $\gets$ \textsc{Eval\_Entropy}(results)
    \If{entropy\_child $>$ entropy\_parent \textbf{and} depth $<$ \textit{d}}
        \State subcells $\gets$ \textsc{Divide\_Cell}(cell, use\_sensitivity)
        \ForAll{subcell in subcells} \textbf{parallel}
            \State \textsc{Explore\_Cell}(subcell, depth + 1, entropy\_child)
        \EndFor
    \EndIf
\EndFunction

\Function{Eval\_Entropy}{samples}
    \State p\_stable $\gets$ fraction of samples labeled stable
    \State p\_unstable $\gets$ 1 - p\_stable
    \State \Return $-(\text{p\_stable} \log_2 \text{p\_stable} + \text{p\_unstable} \log_2 \text{p\_unstable})$
\EndFunction

\Function{Divide\_Cell}{cell, use\_sensitivity}
    \If{use\_sensitivity}
        \State Divide along the most sensitive dimension(s)
    \Else
        \State Divide along preconfigured grid dimensions
    \EndIf
    \State \Return list of subcells
\EndFunction
\end{algorithmic}
\end{algorithm*}

\subsubsection{Experiment 1: Reducing communication overhead}

In this experiment, we assess how rolling windows and aggregation methods impact data transmission in the HP2C-DT framework. The experiment varies two parameters: the sampling interval $T_s$ and the aggregation interval $T_a$. The sampling interval defines how often the hardware collects measurements and sends them to the Edge layer. The aggregation interval determines how frequently the Edge layer transmits data to the Cloud layer. Additionally, for this experiment, we make the aggregation interval define the rolling window size. For example, if the sampling interval is 1 ms and the aggregation interval is 10 ms, each message sent to the cloud node is derived from a rolling window of the last 10 measurements.

We perform the experiment using a single edge node deployed locally and a cloud node running on an on-premises server at BSC. We assign the edge node a single sensor that generates measurements at predefined acquisition intervals of 1 ms, 10 ms, 100 ms, and 1 s. Each acquisition interval is tested with aggregation intervals of 1 ms, 10 ms, 100 ms, 1 s, and 10 s. The case where both intervals are set to 1 ms represents the baseline, where every measurement is sent individually. The cloud node receives these messages and measures the data transmission rate in bytes per second, including message headers of the RabbitMQ middleware.

Two aggregation strategies are tested. The first one is the \texttt{all} aggregation, which groups all measured values without reducing the number of data points. The second is the \texttt{phasor} aggregation, which compresses a rolling window into a single amplitude and angle value. Table~\ref{tab:exp1-all} presents the results for \texttt{all} aggregation, while Table~\ref{tab:exp1-phasor} shows the results for \texttt{phasor} aggregation. Each table displays the measured required bandwidth for different acquisition-aggregation interval pairs, where rows correspond to sampling intervals, and columns to aggregation intervals. The results are the average of five separate executions for each combination of parameters. Note that Table~\ref{tab:exp1-all} includes cases where $T_s \leq T_a$, covering the limit case in which every measurement is sent as a separate message. In contrast, Table~\ref{tab:exp1-phasor} only includes cases where $T_s < T_a$, as the aggregation method requires a minimum number of measurements to operate.

The results for \texttt{all} aggregation (Table~\ref{tab:exp1-all}) show that increasing the aggregation interval consistently lowers the required bandwidth. However, the rate of improvement diminishes: in the first row (1 ms acquisition interval), increasing the aggregation interval from 1 ms to 10 ms reduces bandwidth by more than 5x, but going from 10 ms to 100 ms only reduces it by about 1.7x. This happens because the total number of measurement data remains constant across each row, so the only reduction in required bandwidth comes from decreasing message overhead. The overhead effect is strongest for the smallest aggregation intervals, as the number of messages per second decreases exponentially from left to right. Moving diagonally in the table reveals the true effect of reducing both the number of messages and the number of data points per message, leading to a reduction in required bandwidth proportional to the sampling interval.

\begin{table*}[t!]
    \centering
    \begin{minipage}{0.45\linewidth}
        \centering
        \caption{Required bandwidth (bytes/s) for \texttt{all} aggregation method.}
        \label{tab:exp1-all}
        \begin{tabular}{@{}r@{\hspace{1em}}|r@{\hspace{1em}}r@{\hspace{1em}}r@{\hspace{1em}}r@{\hspace{1em}}r@{}}
            \toprule
            \multirow{3}{*}{\makecell[cc]{Sampling \\ Interval, \\ $T_s$ (ms)}} &  &  &  &  & \\
            & \multicolumn{5}{c}{Aggregation Interval, $T_a$ (ms)} \\
            & 1 & 10 & 100 & 1000 & 10000 \\
            \midrule
            1 & 473009 & 90500 & 52250 & 48425 & 48043 \\
            10 & & 47299 & 9050 & 5225 & 4843 \\
            100 & & & 4730 & 905 & 523 \\
            1000 & & & & 473 & 91 \\
            10000 & & & & & 47 \\
            \bottomrule
        \end{tabular}
    \end{minipage}
    \hspace{0.04\linewidth} 
    \begin{minipage}{0.45\linewidth}
        \centering
        \caption{Required bandwidth (bytes/s) for \texttt{phasor} aggregation method.}
        \label{tab:exp1-phasor}
        \begin{tabular}{@{}r@{\hspace{1em}}|r@{\hspace{1em}}r@{\hspace{1em}}r@{\hspace{1em}}r@{}}
            \toprule
            \multirow{3}{*}{\makecell[cc]{Sampling \\ Interval, \\ $T_s$ (ms)}} &  &  &  & \\
            & \multicolumn{4}{c}{Aggregation Interval, $T_a$ (ms)} \\
            & 10 & 100 & 1000 & 10000 \\
            \midrule
            1 & 49300 & 4930 & 493 & 49 \\
            10 & & 4930 & 493 & 49 \\
            100 & & & 493 & 49 \\
            1000 & & & & 49 \\
            \bottomrule
        \end{tabular}
    \end{minipage}
\end{table*}

\begin{figure*}[t!]
    \begin{subfigure}[t]{0.5\linewidth}
        \centering
        \includegraphics[width=\linewidth]{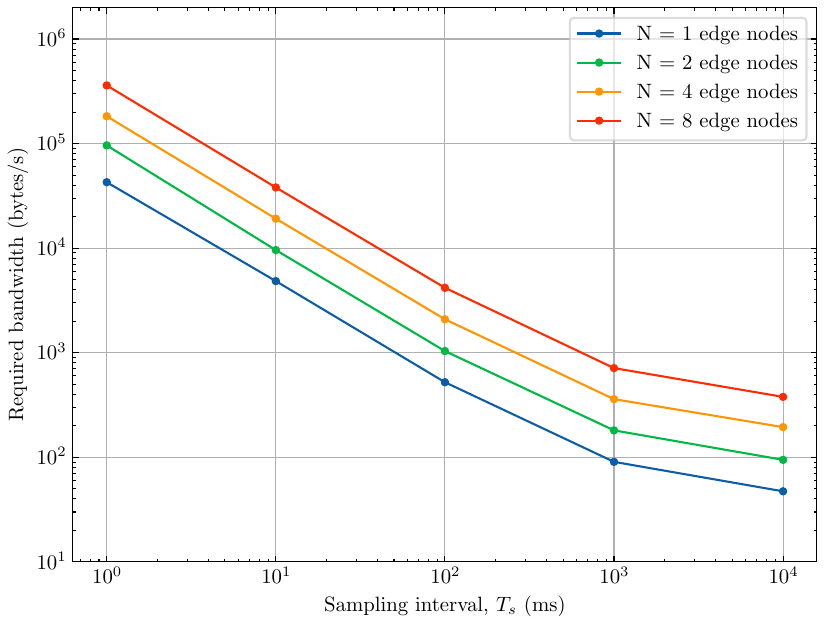}
        \caption{}
    \end{subfigure}%
    \hfill
    \begin{subfigure}[t]{0.485\linewidth}
        \centering
        \includegraphics[width=\linewidth]{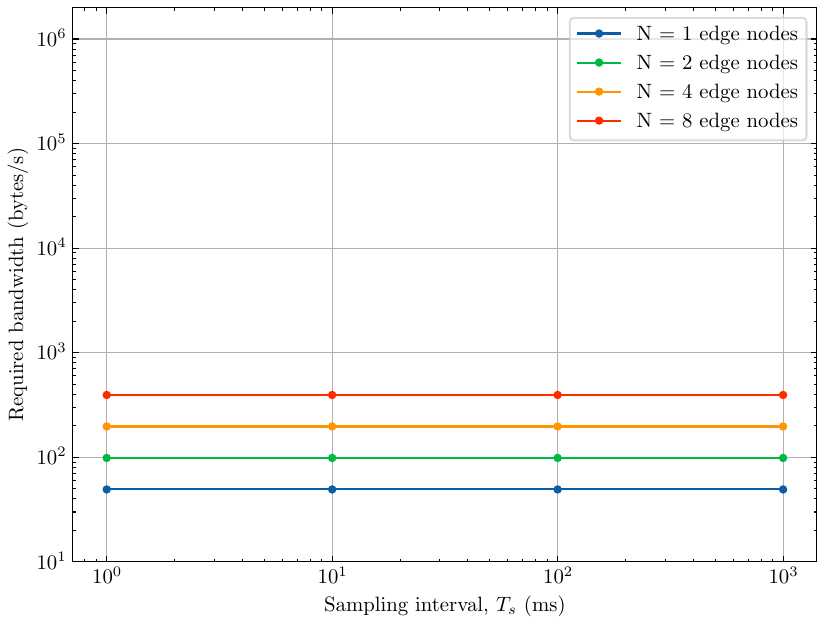}
        \caption{}
    \end{subfigure}
    \caption{Experiment 1: Required bandwidth vs. sampling interval $T_s$ for varying numbers of edge nodes ($N=1,2,4,8$) at fixed aggregation interval $T_a=10$~s. Subfigure (a) shows the \texttt{all} aggregator, where bandwidth decreases with $T_s$ and scales linearly with the number of edges. Subfigure (b) shows the \texttt{phasor} aggregator, where bandwidth remains independent of $T_s$ but scales linearly with the number of edges.}
    \label{fig:exp1-varying-edge-nodes}
\end{figure*}

Table~\ref{tab:exp1-phasor} highlights the impact of actual data aggregation. Since \texttt{phasor} aggregation condenses each rolling window into just two floating-point values (amplitude and angle), the results become independent of the sampling interval. The Required bandwidth decreases proportionally with the aggregation interval, achieving a nearly 10x reduction at each step. In contrast, the \texttt{all} aggregation is limited by the need to transmit every measured value, leading to diminishing improvements.

\begin{figure}[!htbp]
    \centering
    \includegraphics[width=\linewidth]{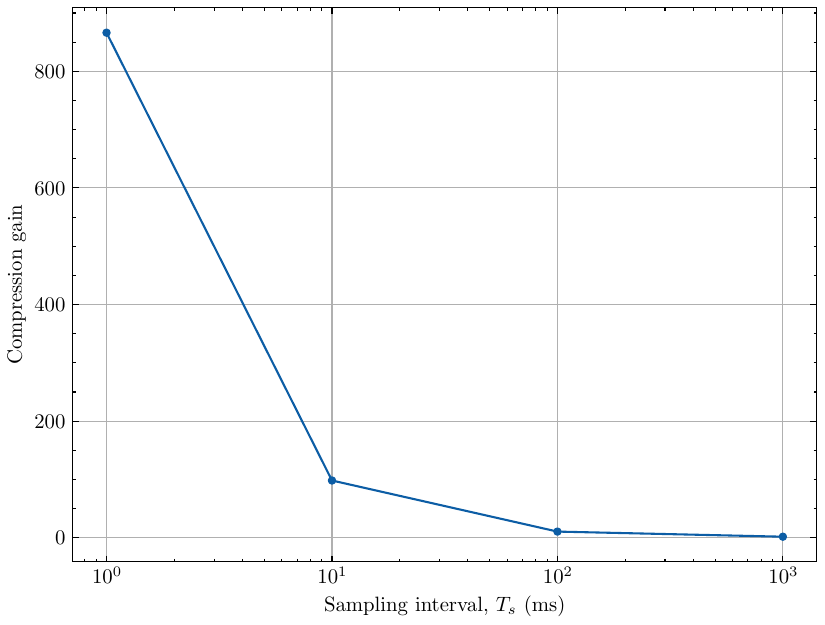}
    \caption{Experiment 1: Compression gain comparison for $T_a = 10$ s.}
    \label{fig:compression-gain}
\end{figure}

Figure~\ref{fig:compression-gain} more explicitly compares the effect of using the two extreme aggregators and shows how the compression gain evolves with the sampling interval. Compression gain is defined as the ratio of bandwidth required by \texttt{all} aggregation to that required by \texttt{phasor} aggregation (see Equation~\ref{eq:compression_gain}). At very fine sampling ($T_s = 1$ ms), the \texttt{phasor} aggregator condenses a large number of measurements per aggregation window, resulting in a massive compression ($\approx$866). As $T_s$ increases, fewer raw measurements fit inside each window, and the benefit decreases to $\approx$1.8 for $T_s = 1000$ ms. Compression is most effective when the ratio $T_s/T_a$ is high and fades as $T_s$ and $T_a$ get closer.

\begin{equation}
    \label{eq:compression_gain}
\text{Compression Gain} = \frac{B_{\mathrm{all}}}{B_{\mathrm{phasor}}}
\end{equation}

We now extend the experiment to multiple edge nodes to assess the impact of scaling on bandwidth requirements. Figure~\ref{fig:exp1-varying-edge-nodes} shows how the required bandwidth scales with the number of edge nodes when varying the sampling interval $T_s$, fixing the aggregation interval at $T_a = 10$ s. This choice ensures the widest coverage along the $x$-axis (since there are no samples for $T_s > T_a$) and also highlights the maximum effect of aggregation, as larger windows provide the strongest bandwidth reduction. Subfigure~(a) confirms that for the \texttt{all} aggregator, the bandwidth decreases with $T_s$ following the same trend across all scales. The four curves are essentially parallel and evenly spaced in logarithmic scale, reflecting a nearly ideal linear scaling with the number of edges at small to moderate scales. Subfigure~(b) shows the case for the \texttt{phasor} aggregator, where the results are independent of $T_s$ and the curves flatten into horizontal lines. As expected, the bandwidth requirements simply scale with the number of edge nodes, but remain constant across different sampling rates owing to the strong compression inherent to the \texttt{phasor} method. Together, these results show that the observed trends in single-node experiments extend to multi-node setups.

This experiment demonstrates the benefits of rolling windows and aggregation in reducing data transmission overhead in the HP2C-DT framework. Rolling windows alone significantly lower the number of messages, reducing the impact of middleware overhead. When combined with aggregation techniques like \texttt{phasor}, which, in turn, takes advantage of domain-specific knowledge, the total amount of transmitted data also decreases, achieving a more efficient data flow while preserving essential information. The HP2C-DT framework integrates these features at the Edge layer, making it adaptable to domain-specific digital twins with varying requirements for latency, bandwidth, and data fidelity. 

While this experiment focuses on measuring bandwidth reduction, it also sets the stage for evaluating trade-offs between communication overhead, data freshness, and accuracy. In this specific setup, both the \texttt{all} and \texttt{phasor} aggregators preserve the signal, so no loss is observed. Still, the framework is flexible and allows domain-specific aggregators and real-world signals, where some information loss may occur. The metrics and approach we used here can easily be extended to study reconstruction error or data age, providing a solid basis for future experiments on these trade-offs in edge–cloud digital twins.

\subsubsection{Experiment 2: Response time for IT/OT actions}

The second experiment evaluates how functions behave in HP2C-DT, focusing on their flexibility in synchronous/asynchronous and distributed execution. Specifically, we measure response times when an action is triggered by a sensor reading. The response time is the duration between the sensor measurement and the execution of a corresponding action on hardware, such as closing a switch, after processing a computational task. We test three execution modes:

\begin{enumerate}
  \item The function runs sequentially on the edge node.
  \item The function executes as a workflow on the edge node using COMPSs.
  \item The function offloads execution as a workflow on a cloud node using COMPSs.
\end{enumerate}

The setup follows the same structure as in Experiment 1. In this case, we configure the system so that the edge node operates with one computing unit for sequential execution (1) and two computing units for workflow execution (2), while the cloud node uses four computing units (3). This setup reflects a real-world edge-cloud environment, where cloud nodes typically have more computational power than edge nodes. The goal is to determine whether offloading computation to the cloud improves performance and under what conditions. It also allows us to assess the benefits of internal parallelization within the edge node.

The experiment examines two sources of overhead: computational overhead from COMPSs task management and communication latency from transmitting execution requests. To assess computational impact, we test functions with different workloads, specifically matrix multiplications of increasing sizes. The computation is parallelized by dividing the multiplication into $m^2$ blocks of size $b^2$, where $m$ is the number of blocks per dimension and $b$ is the block size per dimension. Each time the sensor records a measurement, it triggers a computation task executed through one of the three modes. Once the workflow completes, the function signals a device, the \texttt{MsgAlert} actuator, printing the result and measuring the total time from measurement to result generation.

\begin{figure}[!htbp]
    \centering
    \includegraphics[width=\linewidth]{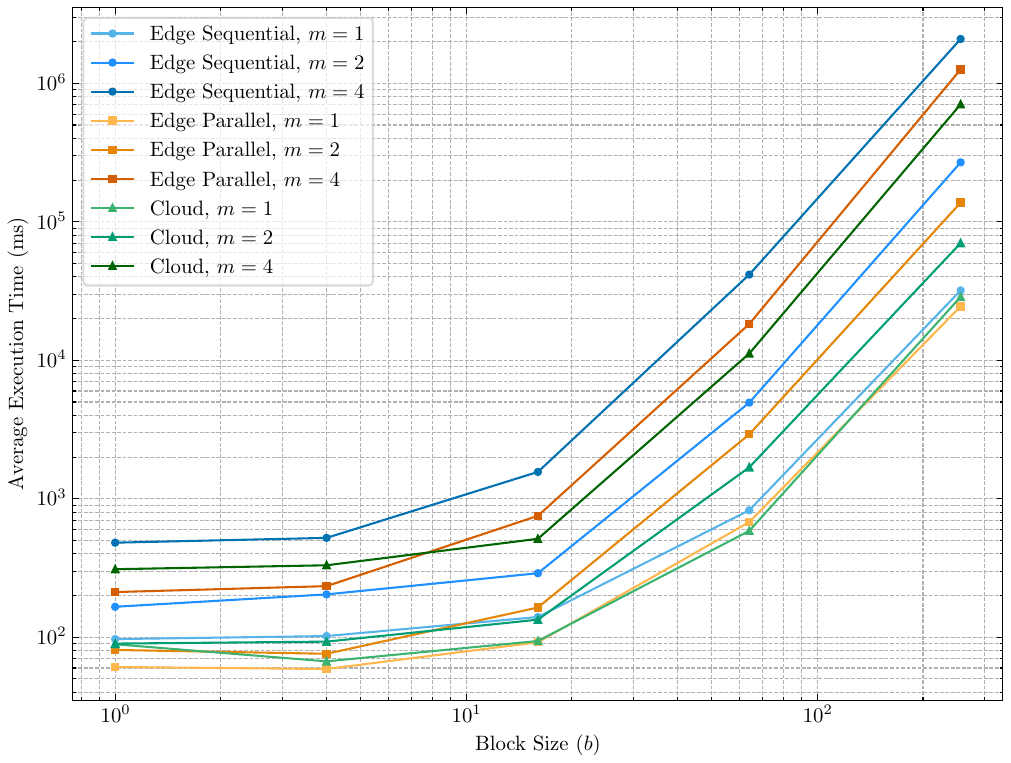}
    \caption{Experiment 2: Response time results.}
    \label{fig:exp2}
\end{figure}

Figure~\ref{fig:exp2} shows the results. For small tasks (lower values of $b$), parallelizing the workflow on the edge node or offloading it to the cloud offers no advantage. As task size increases, there is a break-even point at approximately $b=16$, beyond which offloading reduces response times. Furthermore, as the number of tasks in the workflow increases (higher values of $m$), offloading to the cloud becomes increasingly beneficial. For the configuration with $m=4$ and $b=256$, we observe a 2x speedup in two scenarios: when comparing edge execution (2 CPUs) to sequential execution (1 CPU), and when comparing server execution (4 CPUs) to edge execution. Note the logarithmic scale on the y-axis, so visual differences in execution time do not directly reflect proportional speedups.

Note that the three operation modes shown here are automatically transitioned to by the HP2C-DT framework depending on the needs. The synchronous character of each function is easily defined in the setup files of each node in the network. Plus, when asynchronous, the computing is (a) internally parallelized if possible, and (b) orchestrated and distributed to other resources available seamlessly.

\subsubsection{Experiment 3: Scalability of HPC workloads}

This experiment evaluates how well HPC resources scale when handling computationally intensive tasks. The experiment explores a variable space that includes the power delivered by the 53 converter, the proportion of grid-following vs. grid-forming operation for each converter, the power consumed by the 91 loads, and the control parameters of each converter. The computations are parallelized and distributed using the COMPSs framework and executed on the MareNostrum 5 supercomputer at BSC.

We conduct a strong scaling test, keeping the problem size constant while progressively increasing the computational resources. The test starts with a single node, which has 112 CPUs. The number of nodes is then doubled, progressing as follows: 1 node (112 CPUs), 2 nodes (224 CPUs), 4 nodes (448 CPUs), 8 nodes (896 CPUs), 16 nodes (1,792 CPUs), 32 nodes (3,584 CPUs), and 64 nodes (7,168 CPUs). The data generator explores the variable space down to a \textit{depth} of 3 levels, with a \textit{branching factor} of 4 children per parent subregion, and analyzes 170 operation points per subregion. This results in a total of 14,450 grid stability tasks, each lasting an average of 23.62 seconds. The exploratory nature of the problem and the subregion analysis make it a non-trivial parallel problem, as it requires careful coordination and cannot simply be treated as an embarrassingly parallel application.

\begin{figure*}[!htbp]
    \begin{subfigure}[t]{0.5\linewidth}
        \centering
        \includegraphics[width=\linewidth]{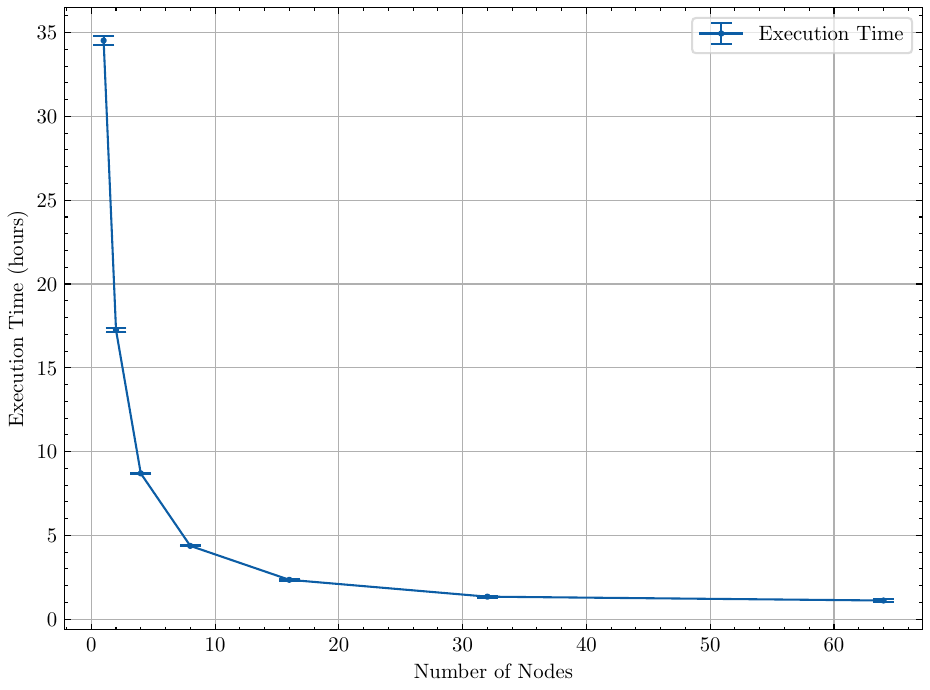}
        \caption{}
    \end{subfigure}%
    \hfill
    \begin{subfigure}[t]{0.485\linewidth}
        \centering
        \includegraphics[width=\linewidth]{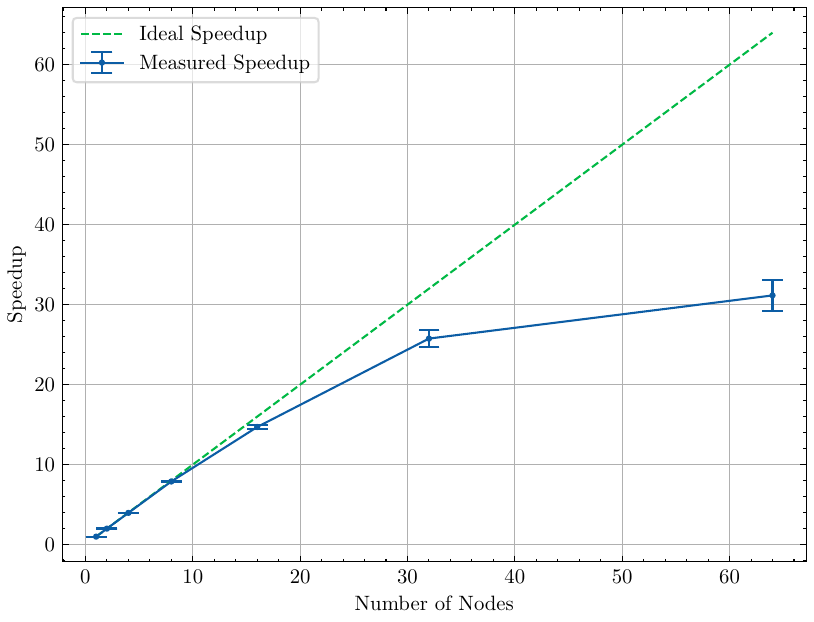}
        \caption{}
    \end{subfigure}
    \caption{Experiment 3: Scalability test results. (a) Average execution time as a function of the number of nodes. (b) Speedup for the same range of cases. Each data point represents the average of 5 runs, with error bars indicating confidence intervals.}
    \label{fig:exp3}
\end{figure*}

Figure~\ref{fig:exp3} presents the scalability results in terms of speedup and execution time. Each configuration is tested with five runs. The results show that speedup remains close to ideal up to 16 nodes. At 32 nodes, the speedup remains acceptable but deviates from the ideal curve, while at 64 nodes, performance saturation is evident. This behavior is expected in strong scaling tests, as increasing the number of nodes eventually leads to higher scheduling overhead. In this case, the need to effectively distribute tasks and aggregate results becomes more challenging, and the benefits of additional parallelism diminish at higher scales.

Beyond raw scalability, the practical impact of HPC resources is significant. With 32 nodes, the execution time is reduced to approximately one hour, compared to multiple hours on a single high-performance server. Even with 16 nodes, the execution takes below two hours. 

In a digital twin scenario, this level of performance enables high-volume offline data generation to be completed within approximately one hour rather than several days. Such accelerated data production allows operators to frequently update or retrain machine learning models for real-time stability assessment, or to refresh stability-related constraints used in real-time optimization. This capability enhances prediction accuracy and system adaptability by keeping models aligned with evolving grid conditions. Achieving this level of computational throughput would be impractical in a purely cloud-based infrastructure due to cost and performance variability.

\section{Discussion}
\label{sec:discussion}

The experiments presented in this work focus on different aspects of the HP2C-DT framework, including data aggregation (Experiment 1), function execution and offloading (Experiment 2), and HPC scalability (Experiment 3). Although a complete system-level benchmark would provide a complete picture of end-to-end performance, such evaluations are highly domain- and infrastructure-dependent and beyond the scope of this study. Still, these experiments provide useful information on the behavior of the framework, help clarify infrastructure needs for domain-specific applications (particularly Experiment 1), and highlight its potential for integrated operation across Edge, Cloud, and HPC layers.

Experiment 1 highlights the advantages of targeted aggregation techniques for reducing communication, while also showing the overhead introduced by the messaging middleware. Regarding messaging overhead, RabbitMQ was chosen not only for raw performance, but for its maturity, reliability, ease of integration with containerized deployments, and native support for publish-subscribe semantics. These properties were important during early architectural development to ensure robustness and flexibility. Rolling-window mechanisms complement this setup, not as a workaround for RabbitMQ’s overhead, but as a strategy to manage bandwidth and scalability across the edge-cloud boundary; something that would likely be needed even with a leaner messaging backend.

Experiment 2 focuses on function execution, comparing sequential processing on the edge, parallel workflows on the edge, and fully or partially offloading to the cloud. The results show that offloading can reduce response times for larger computational tasks and that the framework automatically adapts to available resources while supporting synchronous and asynchronous execution. This illustrates how the architecture can leverage cloud resources without disrupting edge operations.

Experiment 3 evaluates HPC scalability, demonstrating how COMPSs not only supports intra-node parallelization and cloud offloading but also enables substantial speedups for computationally intensive tasks. Even with very large workloads, performance scales well up to 32 nodes (3,584 processors), reducing execution times from several hours to roughly one hour. In the full system, the cloud acts as the main coordination layer, executing decision-making logic, and running mid-tier simulations that complement fast edge-level updates. The HPC cluster is exposed to the cloud as an available resource and is typically used for heavy asynchronous computation campaigns, such as less frequent model updates or broader operational space exploration. This division of roles is what allows the framework to balance responsiveness with computational depth, with the cloud managing the flow of tasks and the HPC layer absorbing the most demanding workloads. Improvements in performance at the HPC layer allow for more frequent updates to digital twin models, which in turn enhances prediction accuracy and system responsiveness of the entire edge-cloud-HPC architecture.

Due to practical reasons, Python is included in HP2C-DT to facilitate the development of digital twins. Its versatility and the ecosystem of AI and data analysis libraries surrounding it make it valuable for modeling and experimentation. Java, on the other hand, smooths the integration of COMPSs and takes care of the time-critical parts of the framework, such as scheduling, orchestration, and hardware interfacing. Java provides runtime-level portability and isolation through the JVM, while Docker containers add system-level packaging (network, file system, process space, dependencies) and make deployment consistent across heterogeneous nodes. This division of roles, together with containerization, ensures that developers can prototype and integrate domain-specific models with minimal friction, without sacrificing execution efficiency where it matters.

Taken together, these observations suggest that HP2C-DT provides a flexible and robust foundation for edge–cloud–HPC digital twin applications. Each subsystem experiment contributes to understanding the behavior of the framework, and the architecture is easily extensible to domain-specific tasks and metrics in future evaluations.

\section{Conclusion}
\label{sec:conclusion}

This article presents a reference architecture that enables seamless coordination between physical devices, cloud nodes, and HPC resources. By distributing tasks based on their requirements---real-time processing at the Edge, global coordination in the Cloud, and intensive computation in HPC---the system balances efficiency and scalability. A key feature is the integration of HPC into the digital twin life cycle, allowing for large-scale probabilistic simulations and optimization tasks, alongside exclusive resource allocation and improved data privacy. At the same time, the architecture retains cloud-based tools for user interaction and global monitoring.

The HP2C-DT framework implements this architecture, not just specifying suitable technologies but also providing a flexible structure to integrate domain-specific components. Powered by COMPSs agents, each node in the network can offload tasks to peer edge/cloud nodes or HPC resources as needed. We test the framework through experiments of communication bandwidth, response time, and computing scalability, which demonstrate the framework’s effectiveness in a power system application.

Future work will focus on decentralizing the architecture by organizing edge nodes into subclusters to improve availability and enable fail-resilient configurations. Additionally, efforts will focus on refining the lifecycle of digital twin models—whether physics-based or data-driven—to ensure their continuous adaptation and deployment across Edge and Cloud layers. This includes further testing how Cloud and HPC resources can be combined to trade Cloud flexibility against HPC queue-based computational power, and integrating adaptive machine learning capabilities that autonomously learn from sensed data, allowing real-time refinement of the digital twin.

\section{Acknowledgments}

This work has been supported by the HP2C-DT TED2021-130351B-C22, HP2C-DT TED2021-130351B-C21 and PID2023-147979NB-C21 projects, funded by the  MCIN/AEI/10.13039/501100011033 and by the European Union NextGenerationEU/ PRTR, and by the Departament de Recerca i Universitats de la Generalitat de Catalunya, research group MPiEDist (2021 SGR 00412). Furthermore, Eduardo Iraola acknowledges his AI4S fellowship within the ``Generación D'' initiative by Red.es, Ministerio para la Transformación Digital y de la Función Pública, for talent attraction (C005/24-ED CV1), funded by NextGenerationEU through PRTR.



\end{document}